\documentclass[]{article}
\usepackage{authblk}
\usepackage[round]{natbib}

\usepackage{amssymb}
\usepackage{amsmath}
\usepackage{amsbsy}
\usepackage{amsthm}
\usepackage{mario-def}

\usepackage{algpseudocode}
\usepackage{algorithm}
\usepackage{graphicx}

\usepackage{url}

\title{Improved design and screening of high bioactivity peptides for drug discovery}
\author[1]{S\'ebastien Gigu\`ere\thanks{sebastien.giguere.8@ulaval.ca}}
\author[1]{Fran\c{c}ois Laviolette}
\author[1]{Mario Marchand}
\author[2]{Denise Tremblay}
\author[2]{Sylvain Moineau}
\author[3]{\'Eric Biron}
\author[4]{Jacques Corbeil}
\affil[1]{Department of Computer Science and Software Engineering, Universit\'e Laval, Qu\'ebec, Canada}
\affil[2]{Department of Biochemistry, Microbiology and Bioinformatics, Universit\'e Laval, Qu\'ebec, Canada}
\affil[3]{Faculty of Pharmacy, Universit\'e Laval, Qu\'ebec, Canada}
\affil[4]{Department of Molecular Medicine, Universit\'e Laval, Qu\'ebec, Canada}

\begin{document}

\maketitle

\begin{abstract}
The discovery of peptides having high biological activity is very challenging mainly because there is an enormous diversity of compounds and only a minority have the desired properties.
To lower cost and reduce the time to obtain promising compounds, machine learning approaches can greatly assist in the process and even replace expensive laboratory experiments by learning a predictor with existing data.
Unfortunately, selecting ligands having the greatest predicted bioactivity requires a prohibitive amount of computational time.
For this combinatorial problem, heuristics and stochastic optimization methods are not guaranteed to find adequate compounds.

We propose an efficient algorithm based on De Bruijn graphs, guaranteed to find the peptides of maximal predicted bioactivity.
We demonstrate how this algorithm can be part of an iterative combinatorial chemistry procedure to speed up the discovery and the validation of peptide leads.
Moreover, the proposed approach does not require the use of known ligands for the target protein since it can leverage recent multi-target machine learning predictors where ligands for similar targets can serve as initial training data.
Finally, we validated the proposed approach \emph{in vitro} with the discovery of new cationic anti-microbial peptides.

Source code is freely available at \url{http://graal.ift.ulaval.ca/peptide-design/}.
\end{abstract}

\section{Introduction}
Drug discovery faces important challenges in terms of cost, complexity and in the rapid discovery of promising compounds.
A valuable drug precursor must have high affinity with the target protein while minimizing interactions with other proteins, in order to avoid side effects.
Unfortunately, only a few compounds have such properties and these have to be identified from an astronomical number of candidate compounds.
Other factors, such as bio-availability, stability, among many others have to be considered; but this combinatorial search problem by itself is very challenging \citep{mee1997design}.
The process of discovering novel compounds with both high bioactivity and low toxicity must therefore be optimized.

Fortunately, in  the last decade, machine learning and kernel methods \citep{ss-02,sc-04} have been extremely effective at providing efficient learning algorithms for a wide range of application domains.
These methods provide novel way to find patterns in biological and chemical data.
These include the prediction of mutagenicity, toxicity and the anti-cancer activity of small molecules~\citep{Swamidass2005Kernels}.
They are also used in the prediction of protein-protein interactions~\citep{BenHur2005Kernel}, protein-ligand interactions and the \emph{in silico} screening of novel targets~\citep{Jacob2008Virtual}.
This success can be mainly attributed to the inclusion of similarity functions, known as \emph{kernels}~\citep{sc-04,ss-02}.
The kernels incorporate valuable biological and chemical knowledge proposed by biologists and, consequently, provide a natural and efficient way to improve the accuracy of learning algorithms.
However,  the use of state-of-the-art learning algorithms to design and enhance the pharmaceutical properties of known compounds have remained largely unexplored and unexploited \citep{schneider2010virtual, damborsky2009computational}.

In the context of drug design, knowing if a ligand will interact with a particular protein is helpful.
However, most potential ligands have low activity and would not represent valid drug precursors.
To predict more valuable information, recent work has addressed the task of predicting the bioactivity and binding affinity between ligands and a target protein \citep{giguere2013learning}.
For instance, starting with a training set containing approximately $50 - 100$ peptides with their corresponding quantified activity (bioactivity, binding affinity, etc) one can expect that a state-of-the-art kernel method will give a bioactivity predictor which is accurate enough to find additional peptides with higher activity than the best ones in the original .
This is possible, since each peptide that possesses a small binding affinity contains information about subsequences of residues that can bind to the target.

For novel and less studied targets, screening libraries remain the method of choice for rapid ligand development.
To fully exploit the great conformational and functional diversity that is accessible with peptides, combinatorial chemistry is certainly the most powerful tool.
A major asset of combinatorial peptide libraries over classic combinatorial libraries, where the scaffold is fixed, is the possibility to generate enormous conformational diversity as well as functional diversity using a randomized synthesis procedure.
This chemical diversity and functionality can be further enhanced by the inclusion of non-natural amino acids.
Furthermore, having a peptide scaffold can be very informative to screen for similarities in peptidomimetics libraries.
However, it is important to note that combinatorial peptide chemistry cannot cover a significant part of the peptide diversity when using more than a few amino acids.
For example, $2$g of a one-bead one-compound combinatorial library composed of randomly-generated peptides of nine residues will generate a maximum of six million compounds, representing a vanishingly small fraction (less than $0.0016\%$) of the set of all $20^{9}$ peptides. 
Consequently, it is almost certain that the best peptides will not be present in the initial screening and most synthesized peptides will have low bioactivity.

The drug discovery challenge is a complex combinatorial problem which unfortunately cannot be solved using combinatorial chemistry alone.
Many have proposed to use combinatorial chemistry or existing databases to learn a machine learning predictor to tackle this problem.
The main motivation was that, \emph{in silico} prediction is fast and inexpensive and would ultimately accelerate this costly process.
Unfortunately, this effectively transforms the combinatorial drug discovery problem into a equally hard optimization task which is sometime know to be NP-Hard \citep{cortes2005general}.
Indeed, predicting the bioactivity of all possible ligands, then selecting the most promising ones, would require a prohibitive amount of computational time.
Heuristics and stochastic optimization are generally the methods of choice when facing such tasks \citep{jamois2003reagent, pickett2000enhancing}.
However, these time and resource consuming methods are not guaranteed to find the optimal solution.
In fact, because of the size of the search space, stochastic optimization is likely to find a poor solution.
The problem of finding the most active and specific ligand is still an open algorithmic problem.

We propose, for a large class of machine learning predictors, an efficient algorithm based on De Bruijn graphs to find the peptide of maximal predicted bioactivity.
This algorithm can be part of an iterative combinatorial chemistry procedure that could speed up the discovery and the validation of peptide leads.
Moreover, the proposed approach can be executed without known ligands for the target protein, since it can leverage recent multi-target machine learning predictors where ligands for similar targets can serve as an initial training set.
Finally, we demonstrate the effectiveness and validate our approach \emph{in vitro} by providing an example of how anti-microbial peptides with proven activity were selected.

\section{Approach}
\subsection{The Generic String kernel}\label{section:GS}
String kernels are similarity functions between strings, which are arbitrary sequences of characters.
In our context, strings are composed of amino acids or nucleotides.
Such kernels have been widely used in applications of machine learning to biology.
For example, the local-alignment kernel \citep{Saigo2004}, closely related to the well known Smith-Waterman alignment algorithm, was used for protein homology detection.
It was observed that kernels for large molecules such as proteins were not suitable for smaller amino acids sequences such as peptides.
Indeed the idea of gaps in the local-alignment kernel or the Smith-Waterman algorithm is well suited for protein homology, but a gap of only a few amino acids in a peptide would have important consequence on the binding affinity with a target protein.

Many recently proposed string kernels have emerged from the original idea of the spectrum kernel \citep{Leslie2002Spectrum} were each string is represented by the set of all $k$-mers that are present. For example, $PALI$ can be represented by it's set of $2$-mers $\{ PA, AL, LI\}$.
As defined by the spectrum kernel, the similarity score between two strings is simply the number of $k$-mers that they have in common.
For example, the spectrum similarity between $PALI$ and $LIPAT$ would be $2$, because they have two $2$-mers in common ($PA$ and $LI$).

To characterize the similarity between peptides, two different $k$-mer criteria were found to be important. First, two $k$-mers should only contribute to the similarity if they are in similar positions in the two peptides \citep{Meinicke2004Oligo}. Second, the two $k$-mers should share common physico-chemical properties \citep{twkr-10}.

\citet{Meinicke2004Oligo} proposed to weight the contribution of identical $k$-mers with a term that decays exponentially rapidly with the distance between their positions. If $i$ and $j$ denote the positions of the $k$-mers in their respective strings, the contribution to the similarity is given by
\begin{equation}\label{eq:position_penalization}
e^{\left(\frac{-(i - j)^2}{2\sigma_p^2}\right)}\, ,
\end{equation}
where $\sigma_p$ is a parameter that controls the length of the decay.

\citet{twkr-10} proposed to consider properties of amino acids when comparing similar $k$-mers.
This was motivated by the fact that amino acids with similar physico-chemical properties can be substituted in a peptide while maintaining the binding characteristics.
To capture the physicochemical properties of amino acids, they proposed to use an encoding function $\psib:\Sigma\longrightarrow \mathbb{R}^d$ where $\psib\mbox{\small $(a)$}=(\psi_1\mbox{\small $(a)$},\psi_2\mbox{\small $(a)$},\ldots\psi_d\mbox{\small $(a)$})$, to map every amino acids $a\in \Sigma$ to a vector where each component $\psi_i(a)$ encodes one of the $d$ properties of amino acid $a$.
In a similar way, we can define $\psib^k: \Sigma^k\longrightarrow\mathbb{R}^{dk}$ as an encoding function for $k$-mers, where 
\begin{equation}\label{eq:psibl}
\psib^k(a_1,a_2,..,a_k) \eqdef (\psib(a_1), \psib(a_2), \ldots, \psib(a_k))\, ,
\end{equation}
by concatenning $k$ physico-chemical property vectors, each having $d$ components.
It is now possible to weight the contribution of two $k$-mers $\ab$ and $\ab'$ according to their properties:
\begin{equation}\label{eq:pc_penalization}
e^{\left(\frac{-\parallel \psib^k(a_1,..,a_k) \,-\, \psib^k(a'_1,..,a'_k) \parallel^2}{2\sigma_c^2}\right)}\, ,
\end{equation}
where $\|\cdot\|$ denotes the Euclidean distance.

More recently, the Generic String (GS) kernel was proposed for small biological sequences and pseudo-sequences of binding interface \citep{giguere2013learning}.
The GS kernel is defined by
\begin{equation}\label{eq:GS_kernel}
\begin{split}
GS&(\xb,\xb', k, \sigma_p, \sigma_c) \eqdef \sum_{l=1}^{k}\sum_{i=0}^{|x|-l} \sum_{j=0}^{|x'|-l}\,  e^{\left(\mbox{\large $\frac{-(i - j)^2}{2\sigma_p^2}$}\right)} \\
&\times e^{\left(\mbox{\large $\frac{-\parallel \psib^l(x_{i+1},..,x_{i+l}) \,-\, \psib^l(x'_{j+1},..,x_{j+l}) \parallel^2}{2\sigma_c^2}$}\right)}\, .
\end{split}
\end{equation}
Hence, the similarity between string $\xb$ and $\xb'$, as defined by the GS kernel, is given by comparing their $1$-mers, $2$-mers, \ldots up to their $k$-mers with the position penalizing term of Equation \eqref{eq:position_penalization} and the physico-chemical contribution term of Equation \eqref{eq:pc_penalization}.
$k$, $\sigma_p$, $\sigma_c$ are hyper-parameters used for tuning the GS kernel and are generally chosen by cross-validation.

This Generic String kernel is very versatile since, depending on the chosen hyper-parameters, it can be specialized to eight known kernels \citep{giguere2013learning}: the Hamming kernel, the Dirac delta, the Blended Spectrum \citep{sc-04}, the Radial Basis Function (RBF), the Blended Spectrum RBF \citep{twkr-10}, the Oligo \citep{Meinicke2004Oligo}, the Weighted degree \citep{RS2004}, and the Weighted degree RBF \citep{twkr-10}.

In a recent study \citep{giguere2013learning}, the GS kernel was used to learn a universal peptide-protein binding affinity predictor capable of predicting, with reasonable accuracy, the binding affinity of any peptide to any protein using the PepX database as the training set.
The GS kernel has also outperformed the current state-of-the-art methods for predicting peptide-protein binding affinities on single-target and pan-specific Major Histocompatibility Complex class II benchmark datasets and three Quantitative Structure Affinity Model benchmark datasets.
Recently, the GS kernel won the 2012 Machine Learning Competition in Immunology \citep{mhcnp2013}.
External validation determined that an SVM classifier equipped with the GS kernel was the overall best method to identify, given unpublished experimental data, new peptides naturally processed by the major histocompatibility complex (MHC) Class I pathway.
The proven effectiveness of this kernel made it ideal to tackle the present problem.

\subsection{The machine learning approach}
In a binary classification setting, the learning task is to predict whether an example has a specific property such as binding to a target molecule.
In this case, the training set consists of positive examples, those having the desired property, and negative examples, those who do not.
In a regression setting, the learning task is to predict a real value that quantifies the quality of a peptide, for example, its bioactivity, inhibitory concentration, binding affinity, or bioavailability.
Such values are generally obtained from \emph{in vitro} or \emph{in vivo} experiments.

In this paper, each example will be of the form $((\xb, \yb), e)$, where $\xb$ represents a peptide and $\yb$ the drug target, which is typically a protein (but other biomolecules could be considered).
In the present regression context, $e \in \mathbb R$ is a real number representing the bioactivity of the peptide $\xb$ with the target $\yb$.
In classification, $e \in \{+1, -1 \}$ denotes if $(\xb, \yb)$ has the desired property or not.
Since predicting real values is strictly more general than predicting binary values, we focused on the more general case of real-valued predictors. 

A predictor can be a function $h$ that returns an output $h(\xb,\yb)$ when given any input $(\xb,\yb)$.
In our setting, the output $h(\xb,\yb)$ is a real number that estimates  the ``true'' bioactivity $e$ between $\xb$ and $\yb$.
Such a predictor is said to be multi-target since it's output depends on the ligand $\xb$ and the target $\yb$.
A multi-target predictor is generally obtained by learning from numerous peptides, binding to various proteins, for example, a protein family.
For this reason, it can predict the bioactivity of any peptide with any protein of the family even if some proteins are not present in the training data.

In contrast, a predictor $h_\yb(\xb)$ is said to be \emph{target-specific} when it is dedicated to predict the bioactivity of a specific (fixed) protein $\yb$ and any peptide $\xb$.
A target-specific predictor is obtained by learning only from peptides binding to a specific protein $\yb$.
For this reason, it can only predict the bioactivity of peptides for the target $\yb$.
In this paper, we focus on the more general case of multi-target predictors.

Given a training set $\{((\xb_1, \yb_1),e_1),\ldots, ((\xb_m,\yb_m),e_m) \}$, a large class of learning algorithms, produce multi-target predictors $h$ with the output $h(\xb,\yb)$ on an arbitrary example $(\xb,\yb)$ given by
\begin{equation}\label{eq:multi-task_predictor}
h(\xb, \yb) = \sum_{i=1}^m \alpha_i k_{\Ycal}(\yb, \yb_i) k_{\Xcal}(\xb, \xb_i)\, ,
\end{equation}
where $k_{\Ycal}$ and $k_{\Xcal}$ are, respectively, the similarity functions between proteins and peptides, and $\alpha_i$ is the weight on the $i$-th training example.
The weight vector $\alb \eqdef (\al_1,\ldots,\al_m)$ depend on the learning algorithm used, but many algorithms produce prediction functions given by Equation \eqref{eq:multi-task_predictor}, including the Support Vector Machine, the Support Vector Regression, the Ridge Regression, Gaussian Processes, \ldots
This makes our solution for drug design compatible with these learning algorithms\footnote{Note that all these learning methods require both kernels to be symmetric and positive semi-definite. This is the case for the GS kernel.}.

Since we use the GS kernel, we have
\begin{equation}\label{eq:GS_linear_combin}
h(\xb, \yb) = \sum_{n=1}^m \beta_n(\yb) GS(\xb,\xb_n, k, \sigma_p, \sigma_c)\, ,
\end{equation}
were  $\beta_n(\yb) = \alpha_n k_\Ycal(\yb, \yb_n)$.

\subsection{The combinatorial search problem}
The general motivation for learning a predictor from training data is that once an accurate predictor is obtained, finding druggable peptides would be greatly facilitated.
It is true that replacing expensive laboratory experiments by an \emph{in silico} prediction will reduce cost.
However, peptides having an average bioactivity do not qualify as drug precursors.
Instead, we should focus on identifying the most bioactive compounds.
The computational problem is thus to identify and rank peptides according to a specific biological function.

Let $\Acal$ be the set of all amino acids, and $\Acal^l$ be the set of all possible peptides of length $l$.
Then, finding the peptide $\xb^\star \in \Acal^l$ that, according to $h$, has the maximal bioactivity with $\yb$, amounts at solving
\begin{equation}\label{pre_image_problem}
\xb^\star_\yb = \arg \max_{\xb \in \Acal^l} h(\xb, \yb)\, .
\end{equation}
This pre-image problem is known to be NP-Hard for several kernels~\citep{GV-09}. Since the number of possible peptides, grows exponentially fast with the length $l$ of the peptide,
a brute force algorithm has an intractable complexity of $\mathcal{O}(|\Acal|^l \cdot \mathcal{O}(h))$ where $\mathcal{O}(h)$ is the time complexity of computing the output of the predictor $h$.
Such an algorithm becomes impractical for any long peptides, such as stapled peptides known to have up to $35$ amino acids.

Stochastic methods such as Metropolis-Hasting are often used to optimize such functions $h$.
These methods are time consuming, highly dependent on optimization parameters and have no guarantee on the solution found.

In the next section, we present an efficient algorithm to find the exact solution to the GS kernel pre-image problem.
The time complexity of the proposed algorithm is linearly dependent on $l$, yielding tractable applications for peptides and proteins.

\section{Methods}
\subsection{Finding the peptide of maximal bioactivity}
In this subsection, we assume that we have in hand a prediction function $h(\xb,\yb)$ in the form of Equation \eqref{eq:GS_linear_combin}.
In this case, we will show how the problem of finding the peptide $\xb^\star_\yb \in \Acal^l$ of maximal bioactivity reduces to the problem of finding the longest path in a directed acyclic graph (DAG).
To do so, we will construct a graph with a source and a sink vertex such that for all possible peptides $\xb \in \Acal^l$, there exists only one path, which has length $h(\xb, \yb)$, that goes from the source to the sink. 
If the size of the constructed graph is polynomial in $l$, any algorithm that efficiently solves the longest path problem in a DAG will also efficiently solve the pre-image problem of the GS kernel.

A bipartite graph $G = ((U,V),E)$ is a graph whose vertices can be divided into two disjoint sets $U$ and $V$ such that every edge in $E$ connects a vertex in $U$ to one in $V$.
For any integer $i$, let $G_i = ((U_i,V_i),E_i)$ be the $i$-th De Bruijn directed bipartite graph of some set.
This means that for every sequence $s\in\Acal^k$, there is a vertex in $U_i$ with the tuple $(s,i)$ as label and a vertex in $V_i$ with the tuple $(s,i+1)$ as label.
Moreover, there is a De Bruijn edge $((u,i), (v, i+1))$ from $(u,i) \in U_i$ to $(v,i+1) \in V_i$ if and only if the last $k-1$ amino acids of $u$ are the same as the first $k-1$ amino acids of $v$.
Note that $\forall i \in \mathbb{N}$, directed edges in $G_i$ only go from vertices in $U_i$ to vertices in $V_i$.
There are exactly $|\Acal|$ edges that leave each vertex in $U_i$ and there are exactly $|\Acal|$ edges that point to each vertex in $V_i$. Moreover, for any chosen integer $k$, $|U_i| = |V_i| = |\Acal^{k}|$ and $|E_i| = |\Acal^{k+1}|$.
Consequently, for each sequence in $\Acal^{k+1}$, there is a single edge going from a vertex in $U_i$ to a vertex in $V_i$. 

The union between bipartite graph $G_i = ((U_i,V_i),E_i)$ and bipartite graph $G_{i+1} = ((U_{i+1},V_{i+1}),E_{i+1})$ is defined to be the $3$-partite graph
\begin{equation*}
G_i\cup G_{i+1}\eqdef ((U_i,V_i\cup U_{i+1}, V_{i+1}), E_i\cup E_{i+1})\, ,
\end{equation*}
where vertices with the same label in $V_i\cup U_{i+1}$ are merged.

More generally we define a $n$-partite graph as the consecutive union of $n-1$ bipartite graphs:
\begin{equation}\nonumber
\begin{split}
G^n &\eqdef G_1 \cup\ldots\cup G_{n-1} \\ 
&\eqdef ((U_1, V_1 \cup U_2, \ldots, V_{n-2} \cup U_{n-1}, V_{n-1}), E_1 \cup \ldots \cup E_{n-1})\, .
\end{split}
\end{equation}

Consequently, there is a one-to-one correspondence between each sequence of $\Acal^{k+n-1}$ and each path $u, \ldots, v: u\in U_1, v\in V_{n-1}$ of length $n$ in $G^n$.

Finally, we add to $G^n$ a source vertex, labeled $(\lambda, 0)$, connected to all vertices in $U_1$ and a sink vertex $t$ connected to all vertices in $V_{n-1}$ ($\lambda$ is the empty string). The set of all possible paths going from the source $(\lambda, 0)$ to the sink $t$ give rise to all sequences in $\Acal^{k+n-1}$.

Using the definition of the GS kernel given at Equation \eqref{eq:GS_kernel} and the general class of predictors given by Equation \eqref{eq:GS_linear_combin}, we can rewrite the generic prediction function as
\begin{equation*}
\begin{split}
h(\xb, \yb) = & \sum_{q=1}^m \beta_q(\yb) \sum_{p=1}^k \sum_{i=0}^{|\xb|-p} \sum_{j=0}^{|\xb_q|-p}\,  e^{\left(\frac{-(i - j)^2}{2\sigma_p^2}\right)} \\
& \times  e^{\left(\frac{-\parallel \psib^p(\xb_{[i+1]},..,\xb_{[i+p]}) - \psib^p({\xb_q}_{[j+1]},..,{\xb_q}_{[j+p]}) \parallel^2}{2\sigma_c^2}\right)}\, .
\end{split}
\end{equation*}
For any string $s$ of length $k$ and any $i\in\{1,\ldots,n\}$, we define
\begin{equation}
\begin{split}
W(s, i) \eqdef & \sum_{q=1}^m \beta_q(\yb) \sum_{p=1}^k \sum_{j=0}^{|\xb_q|-p} e^{\left(\frac{-((i-1) - j)^2}{2\sigma_p^2} \right)} \\
& \times e^{\left(\frac{-\parallel \psib^p(s_1,\ldots, s_p) - \psib^p({\xb_q}_{[j+1]},..,{\xb_q}_{[j+p]}) \parallel^2}{2\sigma_c^2}\right)}
\end{split}
\end{equation}
as the weight on edges $((s',i-1), (s, i))$ where $s' \in \Acal^k \cup \lambda$, $s \in \Acal^k$, and $i \in \{1,\ldots, n\} $ i.e. all edges of $G^n$ except edges heading to the sink vertex $t$.
When $k>1$, edges $((s,n), t)$, heading to the sink vertex $t$, are weighted by the function
\begin{equation}
W_t\big((s, i),t\big) = \sum_{j=1}^{k-1} W(s_{j+1} \ldots s_k, n+j) \, ,
\end{equation}
otherwise, $W_t\big((s, i),t\big) = 0$ when $k=1$.\\
For $n=|\xb|-k+1$, we can now write $h(\xb, \yb)$ as
\begin{equation*}
W_t\big((x_{n},..,x_{|\xb|}, n),t\big) + \sum_{i=1}^{n} W(x_{i},..,x_{i+k-1}, i) \, .
\end{equation*}
Therefore, every path from the source to the sink in $G^n$ builds a unique peptide such that the estimated bioactivity of that peptide is given by the length of the path.

The problem of finding the peptide of highest activity thus reduces to the problem of finding the longest path in $G^n$.
Despite being NP-hard in the general case, the longest path problem can be solved by dynamic programming in $\mathcal{O}(|V(G)| + |E(G|)$ for a directed acyclic graph given a topological ordering of it's vertices.
By construction, $G^n$ is clearly acyclic and its vertices can always be topologically ordered by visiting them in the following order: $(\lambda, 0), U_1, \ldots, U_n, V_n, t$.
Since $G^n$ has $(n |\Acal|^k + 2) \in \mathcal{O}(n|\Acal|^k)$ vertices and $(2|\Acal|^k + (n-1)|\Acal|^{k+1}) \in \mathcal{O}(n|\Acal|^{k+1})$ edges, the complexity of the algorithm will thus be dominated by the number of edges.

We propose an algorithm in $\mathcal{O}(n|\Acal|^{k+1}) = \mathcal{O}((l-k+1)|\Acal|^{k+1})$ for solving the pre-image problem of the GS kernel. Recall that $k$ is a constant and $l$ is the length of the peptide we are trying to identify. Thus $n=l-k+1$.
Small values of $k$ are motivated by the fact that $\parallel \psib^k(a_1,..,a_k) - \psib^k(a'_1,..,a'_k) \parallel^2$
is a monotonically increasing function of $k$ such that Equation \eqref{eq:pc_penalization} vanishes exponentially fast as $k$ increases.
Long $k$-mers will have negligible influence on the dot product and the estimated bioactivity, explaining why small values of $k \leq 6 \ll l$ are chosen by cross-validation.
Therefore, the time complexity of the proposed algorithm is orders of magnitude lower than the brute force algorithm which is in $\mathcal{O}(|\Acal|^{l})$ since $k \leq 6 \ll l$ in practice.
The pseudo-code to find the longest path in $G^n$ is given by Algorithm \ref{algo:longest_path} (supplementary material).

\subsection{Ranking peptides by bioactivity}

Ranking peptides according to their bioactivity will provide valuable information with the potential of accelerating functional peptide discovery.
Indeed, the best peptide candidates can be synthesized quickly by an automated peptide synthesizer and then tested \emph{in vitro}.
Such a procedure will allow quick \emph{in vitro} feedback and minimize turnaround time. 
Also, the best predicted candidates can be utilized to predict a binding motif for a new target protein.
Such a motif should assist researchers in the early study of a target and for the design of peptidomimetic compounds by providing residue preferences.

In the previous section, we have shown how the problem of finding the peptide of greatest bioactivity reduced to the problem of finding a path of maximal length in the De Bruijn graph $G^n$.
By using the same arguments, finding the peptide with the second greatest activity reduces to the problem of finding the second longest path in $G^n$. By induction, it follows that the problem of finding the $K$ peptides of maximal activity reduces to the problem of finding the $K$-longest paths in $G^n$.

Unfortunately, this problem has not been studied much since the longest path problem is generally NP-Complete.
The closely related $K$-shortest paths problem was first studied in $1957$, but attracted most attention following the work of \citet{yen1989efficient, yen1971finding}.
Yen's algorithm was later improved by \citet{lawler1972procedure}. Both algorithms are relatively simple as they use existing shortest path algorithms such as Dijkstra's algorithm to solve the $K$-shortest paths problem.

By exploiting some restrictive properties of $G^n$, we show how Yen's algorithm for the $K$-shortest paths can be adapted to find the $K$-longest paths in $G^n$.
The time complexity of the resulting algorithm, given in Section \ref{sup_mat:k_longest_path} of the supplementary material, is competitive with the latest work on $K$-shortest paths algorithms \citep{eppstein1998finding, brander1995comparative}.

\subsection{From $K$-longest paths to motif}
It is easy to use the $K$-longest paths algorithm to efficiently predicts a motif by simply feeding the $K$ peptides to an existing motif tool.
Here, the motif is a property of the learned model $h(\xb,\yb)$ as opposed to a consensus among known binding sequences.

When the learned model $h(\xb,\yb)$ is a multi-target model, it is then possible to predict affinities for proteins with no known ligand by exploiting similarities with related proteins.
In this case, it is feasible to predict a binding motif for a target with no known binders.
To our knowledge, this has never been attempted successfully.

\subsection{Protocol for split and pool peptide synthesis}
Split and pool combinatorial peptide synthesis is a simple but efficient way to synthesize a very wide spectrum of peptide ligands.
To synthesize several peptides of length $l$ using the $20$ natural amino acids,
the standard approach is to use $20$ reactors for natural amino acids and a pooling reactor.
At every step of the experiment, all reactors are pooled into the pooling reactor which is then split in equal proportions back into the 20 amino acids reactors.
Within this standard approach, each peptide in $\Acal^l$ has an equal probability of being synthesized. 
Since the number of polystyrene beds (used to host every peptide) is generally orders of magnitude smaller than $|\Acal|^l$, there is only a vanishing small fraction of the peptides in  $\Acal^l$ that can be synthesized in each combinatorial experiment. 

Clearly, not every peptide has an equal probability of binding to a target.
More restrictive protocols have been proposed to increase the hit ratio of this combinatorial experiment.
For example, one could fix certain amino acids at specific positions or limit the set of possible amino acids at this positions (for example, only use hydrophobic amino acids).
Such practice will impact the outcome of the combinatorial experiment.
One can probably increase the hit ratio by modifying (wisely) the proportion of amino acids that can be found at different positions in the peptides.
To explore more thoroughly this possibility, let us define a (combinatorial chemistry) \emph{protocol}  $P$ by a $l$-tuple containing, for each position $i$ in the peptide of length $l$, an independent distribution $\mathcal{P}_i(a)$ over the $20$ amino acids $a \in \Acal$.
Hence, we define a protocol by
\begin{equation}\label{eq:probabilistic_protocol}
P \eqdef (\mathcal{P}_1, \ldots, \mathcal{P}_l)\, .
\end{equation} 
Consequently, the peptides produced by this protocol will be distributed following the joint distribution $\mathcal{P}_1 \times \ldots \times \mathcal{P}_l$.
Hence, the probability of synthesizing a peptide is given by
\begin{equation}
P(\xb) = \prod_{i=1}^{|\xb|} \mathcal{P}_i(x_i) \, .
\end{equation}
This family of protocols is easy to implement in the laboratory since, at each step $i$, it only requires splitting the content of the pooling reactor in proportion equal to the distribution $\mathcal{P}_i$ over amino acids.

\subsection{Expected outcome of a library given a protocol}\label{section:protocol_statistics}
We present a method for efficiently computing exact statistics on the outcome of a protocol $P$.
More precisely, we present an algorithm to compute the average bioactivity and its variance over all peptides that a protocol can synthesize.
Note that we cannot compute these statistics by simply predicting the activity of each peptide since the set of all possible peptides of length $l$ is simply too large.  

Such statistics will assist chemists in designing a protocol with a greater hit ratio and avoid superfluous experiments increasing cost.
Indeed, the count of all possible peptides will help to choose the right amount of peptide-coated beads in the assay.
Moreover, the average bioactivity will help designing a protocol that synthesizes as many potential active candidates as possible.
Finally, the bioactivity variance will allow to control the exploration/exploitation trade off of the experiment.

The proposed approach makes use of the graph $G^n$, the protocol $P$, and a dynamic programming algorithm that exploits recurrences in the factorization of first and second order polynomials to efficiently compute the following two quantities:
\begin{eqnarray}
\tau &\eqdef& \sum_{\xb \in \Acal^l} P(\xb) \cdot h(\xb, \yb) \nonumber \\
\beta &\eqdef& \sum_{\xb \in \Acal^l} P(\xb) \cdot h(\xb, \yb)^2 \nonumber\, .
\end{eqnarray}
Indeed, the average and variance bioactivity of peptides synthesized by the protocol are then respectively given by $\tau$ and $\beta - \tau^2$.
The algorithm and its details are given in Section \ref{sup_mat:library_statistics} (supplementary material).

\subsection{Application in combinatorial drug discovery}\label{sec:drug_discovery}

\begin{figure}[!tpb]
\caption{Iterative process for the design of peptide ligand.}
\label{fig:iterativeProcess}
\begin{center}
\includegraphics[scale=0.25]{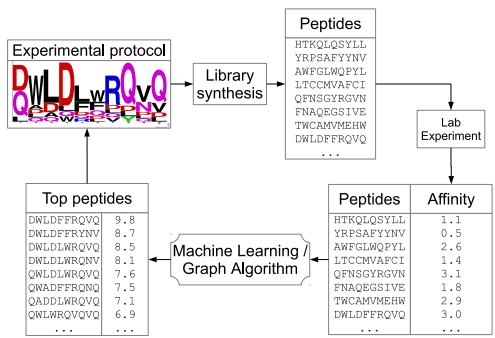}
\end{center}
\end{figure}

We propose an iterative process to accelerate the discovery of bioactive peptide.
The procedure is illustrated in Figure \ref{fig:iterativeProcess}.
First, an initial set of random peptides is synthesized, typically using a split and pool approach.
Then peptides are assayed in laboratory to measure their bioactivities.
At this point most peptides are poor candidates.
They are then used as a training set to produce a predictor $h$. Then $h$ is used for the generation of $K$ bioactive peptides by finding the $K$-longest paths in $G^n$ as described previously.
After, a protocol that consists of using for each position, amino acids with equal proportion in which they appear in the $K$ lead compounds is used to synthesize a large array of potential active peptides.
The algorithm described in Section \ref{section:protocol_statistics} is then used to predict the statistics of the assay.
This ensures that the protocol meets the expectations in terms of quality (average bioactivity) and diversity (bioactivity variance).
To lower the cost, we proceed to synthesize and test the peptides only if the expectations are met.
This process can be repeated until the desired bioactivity is achieved.

\subsection{Peptide synthesis, bacterial strains and minimal inhibitory concentration assay}\label{sec:MIC_assay}
Peptides were synthesized on a Prelude Peptide Synthesizer (Protein Technologies Inc, AZ) using standard Fmoc solid phase peptide chemistry \citep{wellings19974}.
Briefly, the synthesis was performed on Rink Amide AM resin and the amino acid couplings achieved with HCTU/NMM. The peptides were cleaved from the resin using a mixture of $95\%$ trifluoroacetic acid, $5\%$ triisopropylsilane, $5\%$ water for $3$h at room temperature and precipitated in cold diethyl ether.
After triturating for $2$ min, the peptides were collected upon centrifugation and decantation of the ether.
The peptides were purified on a Vydac C18 reversed-phase HPLC column ($22\times250$ mm, $5\mu$m) over $20$ min using a linear gradient of $10-90\%$ acetonitrile with $0.1\%$ trifluoroacetic acid at a flow rate of $10$ mL/min with optical density monitoring at $220$ nm.
The collected fractions were lyophilised and the identity and purity of the peptides assessed by analytical HPLC and MALDI-TOF mass spectrometry.
Peptides were obtained in good yields and with purity greater than $90\%$.

Escherichia coli K12 MG1655 and Staphylococcus aureus 68 (HER1049) were obtained from the F\'elix d'H\'erelle Reference Center for Bacterial Viruses of Universit\'e Laval (\url{www.phage.ulaval.ca}).
Both strains were grown in Trypticase soy broth with agitation at $37^\circ$C. The minimal inhibitory concentration assay was performed as described in \citet{wiegand2008agar}.
The broth microdilution protocol performed in $96$-well plates.
The bacterial strains were grown overnight at $37^\circ$C with aeration and diluted to a final concentration of $5\times 10^5$ cfu/ml in the assay.
The peptides were diluted in sterile water and were tested at the following concentrations: $0, 1, 2, 4, 8, 16$ and $32$ $\mu$g/ml.
The optic density ($600$nm) was followed every $30$ minutes for $24$ hours in a Synergy $2$ plate reader (BioTek Instruments, Inc.).

\section{Result and Discussion}
\subsection{Data}
Two public datasets were used to test and validate our approach.
The first dataset consisted of 101 cationic antimicrobial pentadecapeptides (CAMPs) from the SAPD database \citep{wade2002synthetic}.
Peptide antibacterial activities are expressed as the logarithm bactericidal potency which is the average potency over 24 bacteria such as \emph{Escherichia coli, Bactero\"ides fragilis,} and \emph{Staphylococcus aureus}.
The average antibacterial activity was $0.39$ and the best peptide had an activity of $0.824$.

The second dataset consisted of 31 bradykinin-potentiating pentapeptides (BPPs) reported by \citet{Ufkes1982}.
The bioactivities are expressed as the logarithm of the relative activity index compared to the peptide VESSK.
The average bioactivity was $0.71$ and the best peptide had an activity of $2.73$.

\subsection{Improving the bioactivity of peptides}

To assess the capability of our approach to improve upon known peptides, that is, to predict peptides with superior biological activities to the known ones, we carried out the following experiment using the CAMPs and BPPs peptide datasets.
First, a predictor of biological activity was learned by kernel ridge regression (KRR) for both datasets.
Hyper-parameters for the GS kernel and the KRR were chosen by standard cross-validation.
Then, using the $K$-longest path algorithm and the learned predictor, we generated the $K$ peptides (of the same length as those of the training data) having the greatest predicted biological activity. 

\begin{figure}[!tpb]
\caption{The $K$-longest path algorithm was used to rank the $100,000$ peptides with highest anti-microbial activity. We observed a smooth power law with few exceptional peptides.}
\label{fig:cationic_power_law}
\begin{center}
\includegraphics[scale=0.4]{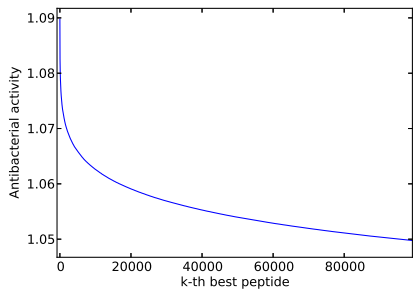}
\end{center}
\end{figure}

On the CAMPs dataset, our approach predicted that the peptide WWKWWKRLRRLFLLV should have an antibacterial potency of $1.09$, a logarithmic improvement of $0.266$ over the best peptide in the training set (GWRLIKKILRVFKGL, $0.824$),
and a substantial improvement over the average potency of that dataset (average of $0.39$).
The anti-microbial activity of the top $100 000$ peptides are showed in Figure \ref{fig:cationic_power_law}.
As expected, we observe a smooth power law with only a few peptides having outstanding biological activity.

On the BPPs dataset, our approach predicted that the pentapeptide IEWAK should have an activity of $2.195$, slightly less than the best peptide of the training set (VEWAK, $2.73$, predicted as $2.192$).
However, the predicted activity of IEWAK is much better than the average peptide activity of the dataset, which is $0.71$.
One can ask why IEWAK has a lower biological activity than VEWAK which was part of the training data?
%Note that only $2$ peptides in the training set have activity greater than $2.0$ and VEWAK has a significantly higher activity.
Machine learning algorithm are known to be resistant to errors and noise in the data.
A possible explanation for this discrepancy is that the biological activity of VEWAK could be slightly erroneous as the learning algorithm could not find a simple predictor given such an outlier.
In addition, VEWAK was predicted an activity of $2.192$ in spite of been seen during training with an activity of $2.73$.
It seems that, given the data, the activity of IEWAK is more easily justified than of VEWAK.

\subsection{Application in drug discovery}

When facing a new target, there is typically none to little information on peptides that could bind to the target protein.
To gather some information, the standard approach is to synthesize a library of random peptides and measure, for each candidate, the desired biological activity.
Typically, this will only yield a few candidates.
The proposed approach has the capability of extracting relevant information from weak candidates and in an iterative manner, predict stronger ones.

As a proof of concept, we propose to replace the laboratory experiment in the proposed procedure by a predictor learned on available datasets.
We will refer to this predictor as the ``expert'', since it encapsulates the current knowledge about the studied problem.
The expert predictor is only used to determine the activities of the peptides generated in the initial random screening phase and those generated by our approach.
To prevent bias, once the expert predictor is learned, the dataset is hidden from the process. 
Then, we randomly draw peptides and, instead of testing them in lab, we use the expert predictor to determine their activities.
At this point most peptides are very weak candidates.
A student predictor is learned with the weak peptides serving as the training data.
We then use the $K$-longest path approach and the student predictor to predict improved compounds.
To validate that peptides predicted (or generated) by this student predictor are indeed relevant, we then validate them with the expert predictor.
Recall that the expert predictor was only used to determine the activity of the randomly generated peptides.

This proof of concept was conducted twice on both the CAMPs and the BPPs datasets. Once by drawing $100$ random peptides, and then by drawing $1000$ random peptides at the initial screening stage.
The results are shown in Table \ref{table:synthetic_results}.
As expected, the number of drawn peptides had no significant effect on the average activity on both datasets.
On both datasets, the number of random peptides had no significant effect on the best peptide found, supporting the development of new techniques to facilitate the discovery of high activity compounds.

Using the proposed approach and the same $100$ random peptides to train the student predictor, we were able to reach an antimicrobial potency of $0.83$, similar to the best peptide of the CAMPs dataset.
By increasing to $1000$ training peptides, we found a peptide having a predicted potency of $1.09$, surpassing the best known peptide.
On the BPPs dataset, the proposed approach considerably outperformed the random peptides method on both the best peptide found and the average bioactivity.
Finally, on both datasets, increasing the number of initial peptides from $100$ to $1000$ had a stronger positive effect than the random approach: $0.26$ increase in potency for CAMPs and $0.16$ increase for BPPs.

\begin{table}
	\center
	\caption{Results from the drug discovery simulation. Comparison between the standard combinatorial screening (random peptides) and the proposed approach (longest paths).}
	\label{table:synthetic_results}
	\begin{tabular}{l|r|rr|rr}
	\multicolumn{2}{c}{} & \multicolumn{2}{c}{\textbf{Random peptides}} & \multicolumn{2}{c}{\textbf{Longest Paths}} \\
	\multicolumn{1}{l}{\textbf{Dataset}} & \multicolumn{1}{r}{\textbf{\# of peptides}} & \multicolumn{1}{r}{\textbf{Average}}  & \multicolumn{1}{r}{\textbf{Max.}} & \multicolumn{1}{r}{\textbf{Average}} & \multicolumn{1}{r}{\textbf{Max.}} \\ 
	\hline
	CAMPs   & $100$ & $-0.58$ & $0.17$ & $0.76$ & $0.83$ \\
			& $1000$ & $-0.59$ & $0.18$ & $1.07$ & $1.09$ \\
	\hline
	BPPs	& $100$ & $0.31$ & $1.39$ & $1.50$ & $2.04$ \\
			& $1000$ & $0.26$ & $1.36$ & $1.66$ & $2.20$ \\
	\hline
	\end{tabular}
\end{table}

\subsection{Binding motifs results}
\begin{figure}[!tpb]
\caption{CAMP bioactivity motifs. \textbf{Top motif:} obtained from the $10^3$ best candidates generated by the expert predictor.  \textbf{Middle motif:} obtained from the $10^3$ best candidates generated by the student predictor. \textbf{Bottom motif:} obtained from the best $10^3$ candidates out of $10^6$ random peptides.}
\label{fig:CAMPs_motifs}
\begin{center}
\includegraphics[scale=0.35]{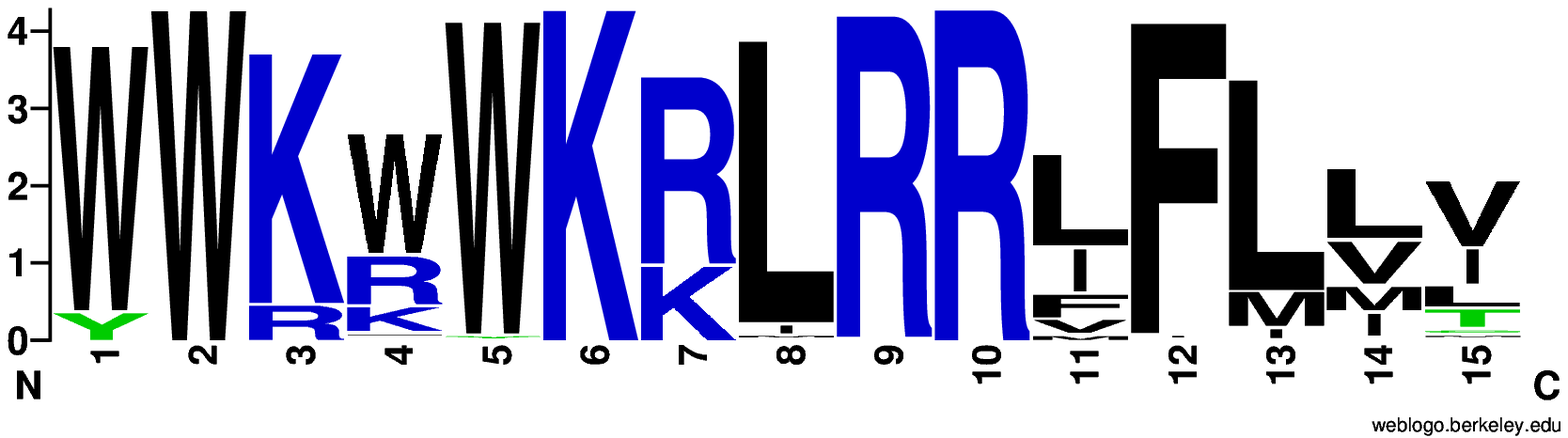}\\
\includegraphics[scale=0.35]{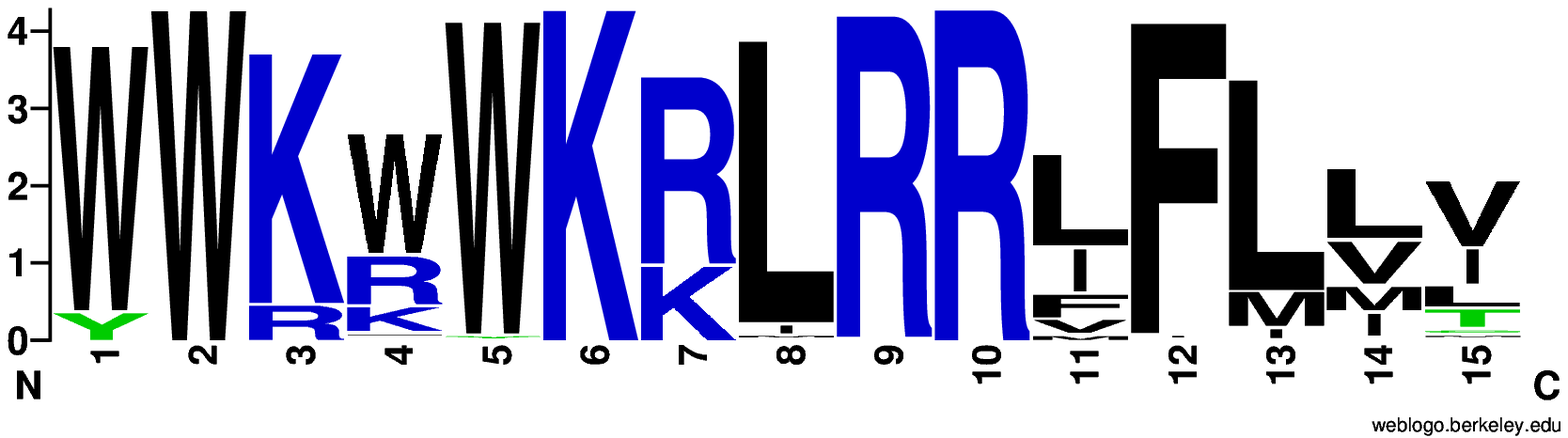}\\
\includegraphics[scale=0.35]{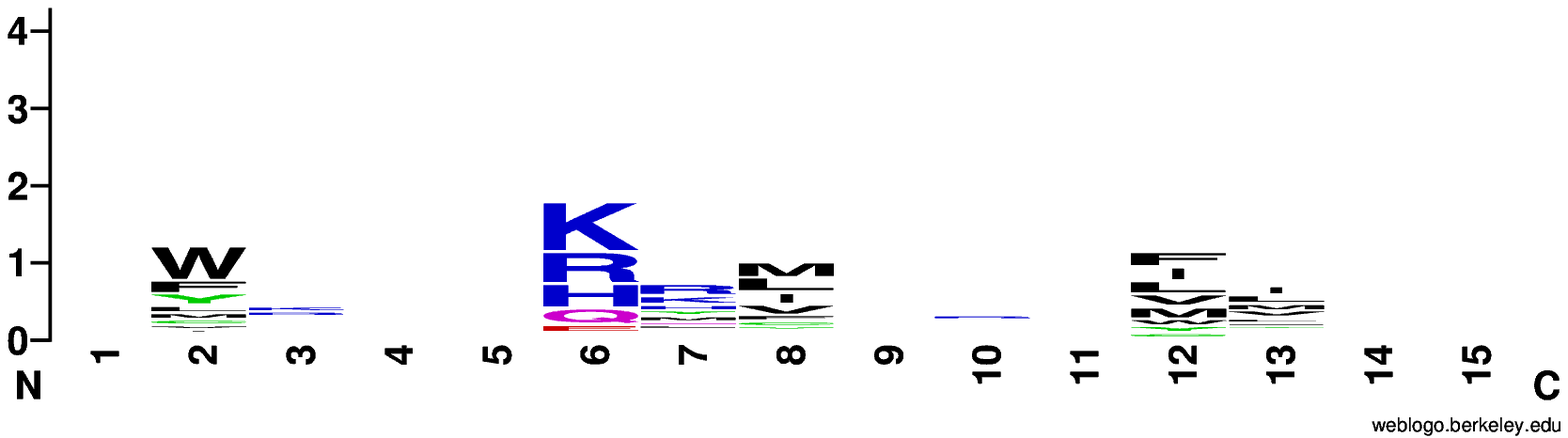}
\end{center}
\end{figure}

To demonstrate the ability of the proposed approach to predict potential functional motifs, we took further the proof of concept proposed in the previous section.
As previously, we learned an expert predictor from the CAMPs dataset which was then hidden for the rest of the procedure.
Using the expert predictor, we predicted the best $10^3$ peptides and produced a bioactivity motif using theses candidates (top panel of Figure \ref{fig:CAMPs_motifs}).
Our goal was to assess how much of that reference motif could we rediscover if we were to hide all of the CAMPs' dataset.
Next, we drew $1000$ random peptides and use the expert predictor to determine their activities.
Theses peptides have, as illustrated in Table~\ref{table:synthetic_results}, on average, very weak antimicrobial potency.
As previously, we learned a student predictor using the random peptides as training data and generated, according to the student predictor, the best $1000$ candidates.
The motif obtained from theses candidates is shown in the middle panel of Figure \ref{fig:CAMPs_motifs}.

We were able to recover all of the reference motif signal using only weakly active peptides.
This provides evidence that the proposed approach could uncover complex signals for new, poorly understood, proteins.
To push the analysis even further, we decreased the number of peptides used to train the student predictor to $100$ training examples.
Even then, for $12$ residue positions, we were able to correctly identify the dominant amino acid property (polar, neutral, basic, acidic, hydrophobic).
This is achieved since the GS kernel encodes amino acids physico-chemical properties.

To put these results in perspective, we took the same peptides used to train the student predictor and generated a motif from them.
The resulting signal was very poor, generating a meaningless motif.
We had to draw $10^6$ random peptides and select the best $10^3$ to produce a motif with minimal information (Figure \ref{fig:CAMPs_motifs} bottom panel).
This clearly illustrates the potential of the proposed approach for accelerating the discovery of potential peptidic effectors.

\subsection{\emph{In vitro} antimicrobial assay}
A total of 12 peptides were synthesized.
The two most active peptides of the CAMPs dataset (Peptide \#$5$ and \#$6$) were used for comparison.
We also selected one peptide with poor activity (Peptide \#$7$) as a control.
The proposed machine learning approach was used to generate a list of $1000$ putative candidates with the highest predicted activity.
From this list, we selected three compounds in a way to maximize the chemical diversity among chosen compounds.
We then tested these peptides (Peptide \#$2$, \#$3$, \#$4$) with a high throughput growth inhibitory assay as described in Section \ref{sec:MIC_assay}.
Results from the minimal inhibitory concentration assay are shown in Table \ref{table:MIC_assay}.
Two of the three candidates have activities equal to the best candidate of the CAMPs dataset.
We were intrigued by the failure of Peptide \#$4$ and after investigation verified that the poor performance was due to poor water solubility.
In a second series, we ensured that a filter for water solubility was employed.
In this second series of tests, Peptide \#$1$ showed (at least against E. coli) better activity than any of the original candidates from the CAMPs dataset, demonstrating that, in this limited biological experiment, we could improve the putative candidates using the proposed machine learning methodology.

\begin{table}
	\center
	\caption{Minimal inhibitory concentration (MIC) resulting from the \emph{in vitro} CAMPs assay. }
	\label{table:MIC_assay}
	\begin{tabular}{c|l|rr}
	\multicolumn{2}{c}{} & \multicolumn{2}{c}{\textbf{MIC ($\mathbf{\mu}g/ml$})} \\
	\multicolumn{1}{c}{\textbf{\#}} & \multicolumn{1}{l}{\textbf{Peptide sequence}} & \multicolumn{1}{r}{\textbf{E. coli}}  & \multicolumn{1}{r}{\textbf{S. aureus}} \\ 
	\hline
	1 &\texttt{YWKKWKKLRRIFMLV} & $2$ & $8$ \\ %Peptide #2 set 2
	2 &\texttt{WWKRWKKLRRIFLML} & $4$ & $4$ \\ %Peptide #3 set 1
	3 &\texttt{WWKRWKRIRRIFMMV} & $4$ & $8$ \\ %Peptide #1 set 2
	4 &\texttt{WWKWWKRLRRLFLLV} & $16$ & $16$ \\ %Peptide #2 set 1
	\hline
	5 &\texttt{KWKLFKGIRAVLKVL} & $4$ & $8$ \\ %Peptide #1 set 1
	6 &\texttt{GWRLIKKILRVFKGL} & $4$ & $4$ \\ %Peptide #6 set 1
	7 &\texttt{KWKLFLGILAVLKVL} & $>32$ & $>32$\\ %Peptide #5 set 1
	\end{tabular}
\end{table}

\section{Conclusion and Outlook}

We proposed an efficient graph-theoretical algorithm that predicts the peptides with the highest biological activities for machine learning predictors using the GS kernel.
We showed how this algorithm can also be used to predict the binding motif for a target with no known ligands.
This is feasible thanks to the multi-target model which is capable of exploiting similarities with related proteins that share common structures.
To increase the hit ratio of combinatorial libraries, we have demonstrated how a combinatorial chemistry protocol relates to a motif distribution.
This allowed us to compute the expected bioactivity and its variance that can be exploited by a combinatorial chemistry protocol such as a one bead one compound protocol.
These steps can be part of an iterative drug discovery process that will have immediate use in both the pharmaceutical industry and academia.
This methodology will reduce costs and the time to obtain lead compounds as well as facilitating their optimization.
Finally, the proposed approach was validated in a real world test for the discovery of new antimicrobial peptides.
These in vitro experiments confirmed the effectiveness of the new compounds uncovered.

The $K$-best peptides were shown to be valuable for the design of split and pool libraries of compounds.
However, in such libraries, it is unclear how we should prioritize high activity candidates (average) over the chemical diversity (variance).
This exploration/exploitation trade-off could be examine in future work. 
Moreover, recent advances in domain-adaptation machine learning could possibly improve the accuracy of the learned predictor since the data generating and the testing distributions are both known.
Finally, the method could be expanded to cyclic peptides and chemical entities (building blocks) of common structure found in clinical compounds.

\section*{Acknowledgement}
JC is the Canada Research Chair in Medical Genomics.

\section*{Funding}
This work was supported in part by the Fonds de recherche du Qu\'ebec - Nature et technologies (FL, MM \& JC; 2013-PR-166708) and the NSERC Discovery Grants (FL; 262067, MM; 122405).

\bibliographystyle{natbib}
\bibliography{arxiv}

\clearpage
\onecolumn

\section{Supplementary material}

\subsection{Algorithm for the longest path}\label{sup_mat:longest_path}
Recall that, by definition, 
\begin{equation*}
\begin{split}
W(s, i) \eqdef & \sum_{q=1}^m \beta_q(\yb) \sum_{p=1}^k \sum_{j=0}^{|\xb_q|-p} e^{\left(\frac{-((i-1) - j)^2}{2\sigma_p^2} \right)} \\
& \times e^{\left(\frac{-\parallel \psib^p(s_1,\ldots, s_p) - \psib^p({\xb_q}_{[j+1]},..,{\xb_q}_{[j+p]}) \parallel^2}{2\sigma_c^2}\right)}\, ,
\end{split}
\end{equation*}
and
\begin{equation*}
W_t\big((s, i),t\big) \eqdef \sum_{j=1}^{k-1} W(s_{j+1} \ldots s_k, n+j) \,\, .
\end{equation*}
The pseudo-code for the longest path algorithm is given by Algorithm~\ref{algo:longest_path}. 

\begin{algorithm}[H]
\caption{Find the longest path in $G^n$ between the source node $(\lambda,0)$ and the sink node $t$.}
\label{algo:longest_path}
\begin{algorithmic}
    \State $length\_to = $ array with $n|\Acal|^k + 2$ entries initialized to $-\infty$
    \State $predecessor = $ array with $n|\Acal|^k + 2$ entries
    \\
    \ForAll{$a\in\Acal^k$}
    	\Comment Edges leaving the source node
         \State $length\_to[a,1] \gets W(a,1)$
    \EndFor
    \\
    \For{$i=2 \to n$}
    	\Comment Edges from the core of $G^n$
        \ForAll{$a\in\Acal^k$}
            \ForAll{$a'\in\Acal$}
                \State $s \gets a_2,\ldots,a_k,a'$
                \Comment Note that $|s| = k$
            	\If{$length\_to[s,i]\leq$$length\_to[a, i-1] + W(s, i)$}
            	\State $length\_to[s,i] \gets length\_to[a, i-1] + W(s, i)$
            	\State $predecessor[s,i] \gets a$
            	\EndIf
            \EndFor
        \EndFor
    \EndFor
    \\
    \State $max\_length \gets -\infty$
    \State $longest\_path \gets \lambda$
    \ForAll{$a\in\Acal^k$}
    	\Comment Edges heading to the sink node
        \If{$max\_length \leq length\_to[a, n] + W_t((a, n), t)$}
            \State $max\_length \gets length\_to[a, n] + W_t((a, n), t)$
            \State $longest\_path \gets a$
        \EndIf
    \EndFor\\
    
    \For{$i=n \to 2$}
    \Comment Backtrack using the predecessors
		\State $a_1, \ldots, a_k \gets predecessor[longest\_path_{[1:k]},i]$
		\State $longest\_path \gets a_1, longest\_path$
    \EndFor\\
\\
\Return $longest\_path$
\end{algorithmic}
\end{algorithm}

\subsection{Algorithm for finding the $K$-longest paths}\label{sup_mat:k_longest_path}
The algorithm uses a trivial variant of the longest path algorithm, given in Section \ref{sup_mat:longest_path}, that allows a path to start from any node of the graph. The pseudo code is given by Algorithm~\ref{algo:k_longest_path}. 

\begin{algorithm}[H]
\caption{Find the $K$-longest paths in $G^n$}
\label{algo:k_longest_path}
\begin{algorithmic}
\State $A = $ array with $K$ entries initialized with the empty string
\State  $B = $ max-heap to store potential paths and their lengths
\State $A[0] \gets$ LongestPath$\big(G^n, (\lambda, 0), t\big)$\\

\For{$i=0 \to K-1$}
	%The spur node ranges from the first node to the next to last node in the shortest path.
	\ForAll{$(a,j) \in \big( (\lambda,0), (A[i]_{[0:k]},1),.., (A[i]_{[l-k:l]},n)\big)$}
	\Comment Nodes of the previous path
%       // Spur node is retrieved from the previous k-shortest path, k ? 1.
%       // The sequence of nodes from the source to the spur node of the previous k-shortest path.
		\State $(V,E) \gets G^n$
		\State root $\gets A[i]_{[0:j+k]}$\\
		
		\For{$r=0 \to i$}
			\If{$A[r]_{[0:j+k]} = root$}
%               // Remove the links that are part of the previous shortest paths which share the same root path.
				\State $E \gets E \setminus (A[r]_{[j:j+k]}, j)$
			\EndIf
		\EndFor\\
		
%       Entire path is made up of the root path and spur path.
		\State $x \gets$ root + LongestPath$\big((V,E), (a,j), t\big)$\\
		
%       // Add the potential k-shortest path to the heap.
		\If{$x \notin B \cup A$}
			\State $B.push\big(x, h(x,y)\big)$
			\Comment Add the string and it's length to the max-heap
		\EndIf
	\EndFor\\
	
	\State $A[i+1] \gets$ B.pop()
	\Comment $B$'s longest path becomes the $i$-th longest path
\EndFor\\
\\
\Return A
\end{algorithmic}
\end{algorithm}

\subsection{Algorithm for computing the combinatorial library statistics }\label{sup_mat:library_statistics}
To compute the statistics efficiently, the dynamic programming algorithm, given by Algorithm~\ref{algo:statistics}, uses the following simple recurrence relations.
\begin{eqnarray}
\sum_{i=1}^n x_i &=& x_n + \sum_{i=1}^{n-1} x_i\, ,
\end{eqnarray}
and
\begin{eqnarray}
%\left( \sum_{i=1}^1 x_i \right)^2 &=& x_1^2 \nonumber \\
\left( \sum_{i=1}^n x_i \right)^2 &=& \left(\sum_{i=1}^{n-1} x_i \right)^2 + 2 x_n \left( \sum_{i=1}^{n-1} x_i \right) + x_n^2 \, .
\end{eqnarray}
Moreover, each node of the graph $G^n$ has the following additional variables.
\begin{itemize}
\item $\taub[s,i]$ for the expected length of paths from the source to the node $(s,i)$.
\item $\btb[s,i]$ for the expected squared length of paths from the source to the node $(s,i)$.
\item $\rhob[s,i]$ for the sum of probabilities given by all possible paths from the source the node $(s,i)$.
\end{itemize}
After executing the dynamic programming algorithm, the values of $\tau$ and $\beta$ are respectively set to $\taub[t]$ and $\btb[t]$ for 
the sink node $t$.
Finally, recall that $\mathcal{P}_i(a)$ is the probability of having amino acid $a$ at position $i$ in a peptide.

\begin{algorithm}[H]
\caption{Compute statistics using $G^n$ and $\mathcal{P}$}
\label{algo:statistics}
\begin{algorithmic}
    \State $\taub, \btb, \rhob : $ arrays with $n|\Acal|^k + 2$ entries initialized to $0$
    \\
    \ForAll{$a\in\Acal^k$}
		\Comment Edges leaving the source node
         \State $\taub[a,1] \gets \mathcal{P}_1(a_1) W((\lambda, 0), (a,1))$
         \State $\btb[a,1] \gets \mathcal{P}_1(a_1) W((\lambda, 0), (a,1))^2$
         \State $\rhob[a,1] \gets \mathcal{P}_1(a_1)$
    \EndFor
    \\
    \For{$i=2 \to n$}
        \ForAll{$a\in\Acal^k$}
            \ForAll{$a'\in\Acal$}
            	\Comment Visiting edge $((a,i-1), (s, i))$
            	\State $s \gets a_2,\ldots, a_k, a'$
				\State $\taub[s,i] += \mathcal{P}_i(a_2) \Big( \taub[a,i-1] + \rhob[a,i-1] W((a,i-1), (s, i)) \Big)$
				\State $\btb[s,i] += \mathcal{P}_i(a_2) \Big( \btb[a,i-1] + \rhob[a,i-1] W((a,i-1), (s, i))^2 + 2\taub[a,i-1]W((a,i-1), (s, i))  \Big)$
				\State $\rhob[s,i] += \mathcal{P}_i(a_2)\rhob[a,i-1]$
            \EndFor
        \EndFor
    \EndFor
    \\
    \ForAll{$a\in\Acal^k$}
    	\Comment Edges heading to the sink node
    	%\State $r \gets \prod_{i=2}^k \mathcal{P}_{n+i}(a_i)$
    	\State $r \gets \prod_{i=1}^{k-1} \mathcal{P}_{n+i}(a_{i+1})$
		\State $\taub[t] += r  \Big( \taub[a,n] + \rhob[a,n] W_t((a, n), t) \Big)$
		\State $\btb[t] += r \Big( \btb[a,n] + \rhob[a,n] W_t((a, n), t)^2 + 2\taub[a,n]W_t((a, n), t)  \Big)$
		\State $\rhob[t] += r \rhob[a,n]$
    \EndFor\\
\\
\Return $\taub[t],\, \btb[t]$ %$\taub[t], \btb[t] - \taub[t]^2$

\end{algorithmic}
\end{algorithm}

\end{document}